\documentclass{article}

\usepackage[english]{babel}

\usepackage[letterpaper,top=2cm,bottom=2cm,left=3cm,right=3cm,marginparwidth=1.75cm]{geometry}

\usepackage{amsmath}
\usepackage{amssymb}
\usepackage{amsthm}
\usepackage{graphicx}
\usepackage[colorlinks=true, allcolors=blue]{hyperref}
\usepackage{booktabs}
\usepackage{comment}

\usepackage{xcolor}
\usepackage{tikz}
\usetikzlibrary{shapes.geometric}
\usepackage{pgfplots}
\usepackage{subcaption}

\newtheorem{remark}{Remark}

\title{Specifying Autonomous System Behaviour\\
\normalsize{Formal Specifications and their Challenges}}

\author{Andrew Sogokon$^1$, Burak Yuksek$^2$,~Gokhan Inalhan$^2$, Neeraj Suri$^1$}
\date{
$^1$~School of Computing and Communications, Lancaster University, UK\\
$^2$~School of Aerospace, Transport and Manufacturing,~Cranfield University, UK\\
\ \\
\today
} 

\begin{document}
\maketitle

\begin{abstract}
Specifying the intended behaviour of autonomous systems is becoming increasingly important but is fraught with many challenges. This technical report provides an overview of existing work on specifications of autonomous systems and places a particular emphasis on \emph{formal specification}, i.e. mathematically rigorous approaches to specification that require an appropriate formalism. Given the breadth of this domain, our coverage is necessarily incomplete but serves to provide a brief introduction to some of the difficulties that specifying autonomous systems entails, as well as existing approaches to addressing these difficulties currently pursued in academia and industry.
\end{abstract}

\section{Introduction}
Autonomous systems are commonly understood to be systems that are capable of decision making independently of their operator; these can be purely discrete computer systems (e.g. software artefacts) or cyber-physical systems (CPS) which operate in a physical environment (e.g autonomous vehicles).
Modern autonomous systems in the CPS domain may employ machine learning components in their control loop and there is a degree of ambiguity about what precisely constitutes a genuine \emph{specification} for an autonomous system (here by a \emph{specification} we shall understand a description of what a system should do). 
In applications where the autonomous system is designed to perform a function that is also commonly performed by human operator, it has previously been suggested (e.g. in~\cite{fisher2013verifying},~\cite{webster2011formal}) that the specification of autonomous system behaviour (or the component responsible for autonomy in a given system) is fundamentally linked to the certification of the human who could otherwise operate the system if it were non-autonomous. In safety-critical applications where malfunction can lead to disastrous outcomes (such as loss of life) it becomes important to specify the desired behaviour precisely and unambiguously, which can be highly non-trivial. Formal logics (e.g. see~\cite{fisher2013verifying}) can be employed to precisely describe the desired system behaviour (such as e.g. the requirement that collisions with obstacles must be avoided throughout the duration of the system's operation) and can provide a certain standard for rigorous specifications.

\begin{remark}
There exist a large number of formalisms for both modelling system behaviour and specifying the desired system properties (a helpful review may be found in the recent survey by Luckcuck et al.~\cite{Fisher-AS-Survey}); these formalisms vary in what properties may be expressed and what systems may be modelled. For example, \emph{temporal properties} such as a system attaining a certain state in the future or a system never colliding with obstacles are examples of properties which can be stated (i.e. formally specified) using \emph{temporal logic} (which we briefly describe in latter sections); however, if one indeed intends to treat an autonomous system \emph{as a human}, one may also wish to reason about the \emph{beliefs} of this system in order to judge whether its actions are justified. The choice of logic in which one wishes to formally express the specification is thus highly significant.
\end{remark}

While formal specifications are standard in areas of computer science that align closely with \emph{formal methods}, they
are presently not very widespread within industrial practice, where carefully worded natural language prose is still commonly employed to specify the intended behaviour of artefacts. The challenges of formalising natural language requirements have long been recognised~\cite{mokos2020survey}, not least because practitioners find logical formalism counter-intuitive (see e.g.~\cite{holzmann2019formalizing}). The most prominent applications of formal system modelling and specification have to date been largely confined to the design of \emph{safety-critical systems} where the cost of malfunction is deemed prohibitive. The natural language requirements for such a system are typically elaborated in a \emph{requirements document} and a formal model can be developed with respect to a formalisation of these requirements (see e.g.~\cite{hoang2013introduction}). The use of formal methods in industrial projects, while uncommon, has seen some notable successes, such as the design of a driverless airport shuttle system~\cite{abrial2006formal}.~\footnote{Notwithstanding these successes, it has been remarked that one of the main problems in applying formal methods such as the B method is the fact that the requirements document is often either badly written or altogether missing \cite{abrial2007formal}.}

The following sections will review some of the existing work and identify challenges in specifying the behaviour of autonomous systems, in particular with regard to imprecise regulations, which autonomous systems are expected to comply with.

\section{Specifications Stemming from Regulation}
\label{sec:regulationSpecs}
As human operators are typically subject to regulations that outline  acceptable behaviour (e.g. the \emph{Highway Code} for road safety and vehicle rules, or the \emph{Rules of the Air} for piloted aircraft), the regulations themselves provide an immediate source of specifications for autonomous systems in which the human operator is dispensed with. Unsurprisingly, existing regulations were designed and written to be interpreted by humans and are therefore stated in natural language. While in theory one may take existing regulation such as the Highway Code/Rules of the Air and replace the word `driver'/`pilot' with `autonomous system', the resulting natural language document will remain imprecise and will constitute an ambiguous set of specifications.

\subsection{Terrestrial Domain}
In the terrestrial domain, e.g. when dealing with autonomous cars, a typical mission simply involves navigating the vehicle from its current position to the desired geographic location; however, the main challenge lies in modelling and validating acceptable road behaviour along the way (as well as reasoning about expected and unexpected behaviour of human drivers)~\cite{kress2021formalizing}.

As in the aerial domain, regulations (traffic laws) exist in the terrestrial domain and provide the bulk of the specifications for the intended behaviour of autonomous cars. While these laws can, in some cases, be well defined and lead to some anticipated safety requirements such as ensuring collision freedom, there is considerable ambiguity as to the interpretation of ``reasonable care'' that drivers are required to exercise in order not to harm other drivers. As remarked in~\cite{shalev2017formal}, any interpretation of this requirements must follow societal norms, \emph{which may be subject to change}. 

\begin{remark}
When certifying human drivers, the driving instructor in a driving test provides his/her \emph{interpretation} of the rules in the Highway Code and acts as a monitor for violations with respect to this interpretation. 
\end{remark}

Shalev-Shwartz et al. at \emph{Mobileye}~\cite{shalev2017formal} attempted to meet this challenge by providing their own unambiguous interpretation of ``reasonable care'' and proposed \emph{responsibility-sensitive safety} (RSS), which is summarised in five ``common sense'' rules:

\begin{itemize}
\item Do not hit someone from behind.
\item Do not cut-in recklessly.
\item Right-of-way is given, not taken.
\item Be careful of areas with limited visibility.
\item If you can avoid an accident without causing another one, you must do it.
\end{itemize}

In order to give an unambiguous interpretation of these rules, a formalisation is undertaken in~\cite{shalev2017formal} which does not employ temporal logic like Fisher's work~\cite{fisher2013verifying}, but takes a more conventional approach in making notions such as e.g. the minimum safe distance between vehicles mathematically precise by phrasing these requirements explicitly in reference to a mathematical model in first-order logic.

\begin{remark}
    The RSS rules in~\cite{shalev2017formal} have been critiqued for their focus on ensuring that the autonomous vehicle does not cause an accident~\cite{torres2020case} (i.e. cannot be blamed for causing one), rather than avoiding accidents altogether.
\end{remark}

\begin{figure}[h!]
    \centering
    \begin{tikzpicture}[thick,scale=0.75, every node/.style={scale=0.75}]


\draw [line width=50.0,gray]  plot[smooth, tension=.7] coordinates {(-6,0) (9,0.0) };
\draw [line width=1.5,white, dashed]  plot[smooth, tension=.7] coordinates {(-6,0.5) (9,0.5) };
\draw [line width=1.5,white, dashed]  plot[smooth, tension=.7] coordinates {(-6,-0.5) (9,-0.5) };

\draw [line width=15.0,red!15, opacity=0.5]  plot[smooth, tension=.7] coordinates {(-6,-1.10) (1.5,0)};
\draw [line width=15.0,blue!15, opacity=0.5]  plot[smooth, tension=.7] coordinates {(-6,-1.10)  (-2,-0.7) (2,1) (6,-1)(9,-1.12) };
\draw [line width=15.0,green!15, opacity=0.5]  plot[smooth, tension=.7] coordinates {(-6,-1.10)  (-2,-0.7) (2,1) (6,1.1)(9,1.12) };

\node[draw, fill=red!85, rounded corners=3pt, minimum width=50pt, minimum height=25pt] at (-5,-1.05) {};
\node[trapezium, trapezium angle=67, fill=black, rotate=90, rounded corners=2pt,  minimum width=3pt, minimum height=12pt] at (-4.8,-1.05) {};
\node[trapezium, inner ysep=2pt, trapezium angle=27, fill=black,  rounded corners=1pt,  minimum width=0.1pt, minimum height=0.1pt] at (-5.2,-1.35) {};
\node[trapezium, inner ysep=2pt, trapezium angle=27, fill=black,  rounded corners=1pt, rotate=180,  minimum width=0.1pt, minimum height=0.1pt] at (-5.2,-0.75) {};
\node[trapezium, trapezium angle=67, fill=black, rotate=-90, rounded corners=2pt,  minimum width=3pt, inner ysep=2pt, inner xsep=6pt, minimum height=5pt] at (-5.6,-1.05) {};
\node[trapezium, trapezium angle=77, fill=yellow, rotate=90, rounded corners=0.1pt,  minimum width=4pt, minimum height=1pt, inner ysep=0.3pt, inner xsep=0.5pt] at (-4.17,-1.33) {};
\node[trapezium, trapezium angle=77, fill=yellow, rotate=90, rounded corners=0.1pt,  minimum width=4pt, minimum height=1pt, inner ysep=0.3pt, inner xsep=0.5pt] at (-4.17,-0.77) {};
\node[trapezium, trapezium angle=70, fill=red, rotate=-70, rounded corners=0.1pt,  minimum width=3pt, minimum height=1pt, inner ysep=0.3pt, inner xsep=0.5pt] at (-4.6,-0.6) {};
\node[trapezium, trapezium angle=70, fill=red, rotate=-120, rounded corners=0.1pt,  minimum width=3pt, minimum height=1pt, inner ysep=0.3pt, inner xsep=0.5pt] at (-4.6,-1.5) {};


\node[draw, fill=red!85, rounded corners=3pt, minimum width=50pt, minimum height=25pt] at (2,-1.2222) {};
\node[trapezium, trapezium angle=67, fill=black, rotate=90, rounded corners=2pt,  minimum width=3pt, minimum height=12pt] at (2.2,-1.2222) {};
\node[trapezium, inner ysep=2pt, trapezium angle=27, fill=black,  rounded corners=1pt,  minimum width=0.1pt, minimum height=0.1pt] at (1.8,-1.5222) {};
\node[trapezium, inner ysep=2pt, trapezium angle=27, fill=black,  rounded corners=1pt, rotate=180,  minimum width=0.1pt, minimum height=0.1pt] at (1.8,-0.9222) {};
\node[trapezium, trapezium angle=67, fill=black, rotate=-90, rounded corners=2pt,  minimum width=3pt, inner ysep=2pt, inner xsep=6pt, minimum height=5pt] at (1.4,-1.2222) {};
\node[trapezium, trapezium angle=77, fill=yellow, rotate=90, rounded corners=0.1pt,  minimum width=4pt, minimum height=1pt, inner ysep=0.3pt, inner xsep=0.5pt] at (2.83,-1.5022) {};
\node[trapezium, trapezium angle=77, fill=yellow, rotate=90, rounded corners=0.1pt,  minimum width=4pt, minimum height=1pt, inner ysep=0.3pt, inner xsep=0.5pt] at (2.83,-0.9422) {};
\node[trapezium, trapezium angle=70, fill=red, rotate=-70, rounded corners=0.1pt,  minimum width=3pt, minimum height=1pt, inner ysep=0.3pt, inner xsep=0.5pt] at (2.4,-0.7722) {};
\node[trapezium, trapezium angle=70, fill=red, rotate=-120, rounded corners=0.1pt,  minimum width=3pt, minimum height=1pt, inner ysep=0.3pt, inner xsep=0.5pt] at (2.4,-1.6722) {};








\end{tikzpicture}
    \caption{Interpreting ``duty of care''/``endangering'' other vehicles}
    \label{fig:my_label}
\end{figure}
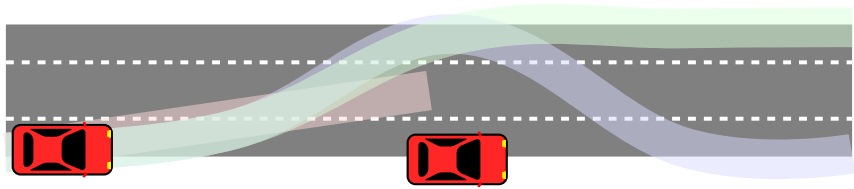

A rather different formalisation effort which also focused on traffic rules for autonomous vehicles is described in the work of Rizaldi et al~\cite{rizaldi2017formalising}, who formalised some (German) traffic rules that govern overtaking vehicles, such as ``\emph{when changing the lane to the left lane during overtaking, no following road user
shall be endangered.}''. In this work, once again, the authors needed to provide a precise meaning of `endangered' in the informal regulations document (a process which the authors termed ``concretisation''). The work in~\cite{rizaldi2017formalising} is noteworthy for employing a theorem proving system (Isabelle) to formalise a subset of traffic rules.

Regulations in the terrestrial domain concern more than just road vehicles. For example, all new railway tracks and rolling stock on European railways are legally reqired to adopt the European Train Control System (ETCS). The specification of ETCS describes how
trains move and periodically receive a Movement Authority. 
Natural language requirements such as 
``\emph{The train trip shall issue
an emergency brake command, which shall not be revoked until the train has reached
standstill and the driver has acknowledged the trip}''. A formalisation of requirements such as these was undertaken by Cimatti et al. in~\cite{cimatti2009requirements}.
Platzer's later work~\cite{platzer2009european} on ETCS formalisation and verification employed a different formalism and a special purpose theorem proving system (KeYmaera~\cite{platzer2008keymaera}).
This system was also employed by Loos et al.~\cite{loos2011adaptive} 
 to formally verify
safe separation between road vehicles.

\begin{remark}
The environment in which autonomous vehicles operate provides a big source of uncertainty when it is not controlled, which in turn makes it difficult to guarantee properties resulting from the interaction of the vehicle with the environment. This fact has long been appreciated, along with the fact that the environment is in practice very difficult to model, which leads one to having to consider abstractions. One common way to address this issue is described by Fremont, Sangiovanni-Vincentelli and Seshia~\cite{fremont2021safety} who argue that in order to specify the behaviour of autonomous vehicles one needs to state assumptions about the environment as part of a \emph{contract}.
\end{remark}

\subsubsection{Notions of Safety}
An important distinction exists between so-called \emph{functional safety} and \emph{nominal safety}~\cite{shalev2017formal}, where functional safety is concerned with hardware and software failures that can lead to a safety hazard (as covered in ISO 26262) and nominal safety concerns itself with the safety of the logical decisions of an autonomous vehicle, assuming that there are no hardware or software malfunctions.
%

Absence of collisions with cars and other obstacles, which is the most common form of specification that has been addressed in existing literature, along with similar requirements such as ensuring safe separation between vehicles~\cite{loos2011adaptive} would fall under the definition of nominal safety.

On the other hand, many industrial requirements fall under the definition of functional safety. Some interesting examples can be found in the recently proposed benchmarks contributed by Eddeland et al.~\cite{eddeland2020industrial}, which feature (modified) safety specifications from the Volvo Car Corporation. Safety requirements in this instance are those properties that must hold at all times.\footnote{The specifications are available on GitHub~\href{https://github.com/decyphir/ARCH20\_ATwSS/tree/master/STLFiles}{https://github.com/decyphir/ARCH20\_ATwSS/tree/master/STLFiles}}

\subsection{Aerial Domain}

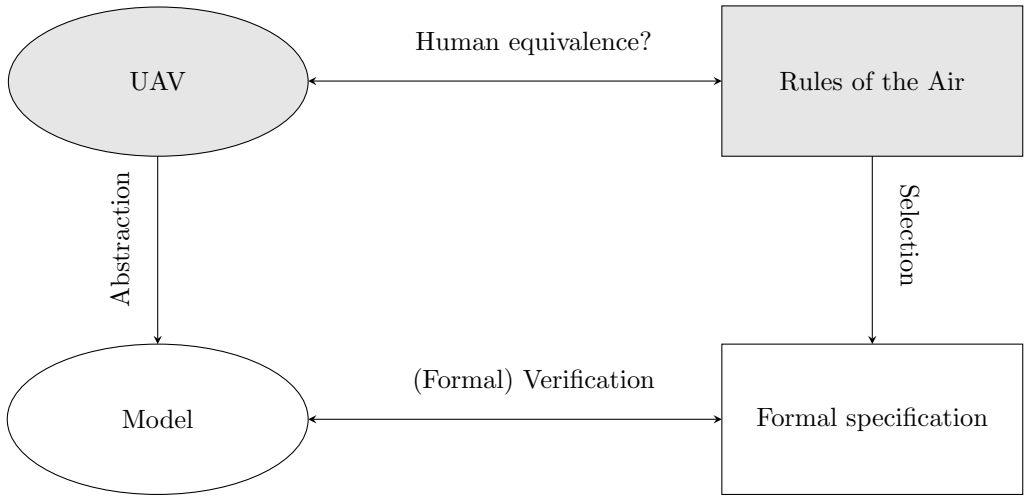
\begin{figure}[ht!]
    \centering
\begin{tikzpicture}
\draw[fill=black!10]  (-3.4925,3.4926) ellipse (1.9925 and 0.9926);
\draw[fill=black!10]   (4,4.5) rectangle (8,2.5);
\draw  (-3.5,-1) ellipse (2 and 1);
\draw  (4,0) rectangle (8,-2);
\node at (6,3.5) {Rules of the Air};
\node at (-3.5,3.5) {UAV};
\node at (-3.5,-1) {Model};
\node at (6,-1) {Formal specification};
\draw[<->, >=stealth] (-1.5,3.5) -- (4,3.5);
\draw[<->, >=stealth] (-1.5,-1) -- (4,-1);
\draw[->, >=stealth] (-3.5,2.5) -- (-3.5,0);
\draw[->, >=stealth] (6,2.5) -- (6,0);
\node at (1.5,4) {Human equivalence?};
\node at (1.5,-0.5) {(Formal) Verification};
\node[rotate=90] at (-4,1.4) {Abstraction};
\node[rotate=-90] at (6.5,1.45) {Selection};
\end{tikzpicture}
    \caption{Certification approach pursued by Webster et al.~\cite{webster2011formal}}
    \label{fig:Webster}
\end{figure}

A notable investigation into the use of formal specification and verification (using an automated verification technology known as~\emph{model checking}) in order to prove that a control system of an autonomous aircraft follows \emph{Rules of the Air} was reported by Webster et al. in~\cite{webster2011formal}. As remarked above, this part of the specification for the autonomous system is in effect  provided by the \emph{Civil Aviation Authority} in the form of a natural language regulations document which describes what behaviour is expected from a pilot. The overall goal of this work was to establish some form of ``human equivalence'' of the UAV with respect to the existing regulations. In order to (partly) achieve this, the authors followed a scheme illustrated in Fig.~\ref{fig:Webster}, where a selection of important regulations is picked out, formalised (i.e. stated unambiguously in a formal logic) and evaluated against a \emph{model} of the UAV, which provides a mathematical abstraction which can be analysed with respect to the formal specification. 
\begin{remark}
The scheme in Fig.\ref{fig:Webster} follows a familiar pattern used in formal modelling and verification where the real world model is abstracted to an appropriate abstract model and the informal requirements are mapped to a formal specification (the interested reader may find a good overview in J. Wing's  article.~\cite{wing1990specifier}).
\end{remark}
The difficulty in specifying the behaviour of autonomous systems such as UAVs subject to Rules of the Air precisely (formally) stems from the fact that certain natural language passages are subject to multiple interpretations. 
The document describing the Rules of the Air Regulations comprises some 15,000 words of natural language text. Formalising the rules in their entirety would not be practically feasible. The work by Webster et al. instead focused on formalising a very small subset of this document, namely the passages below 

\begin{itemize}
    \item \emph{``when two aircraft are approaching head-on and there is danger of collision, each shall alter its course to the right''} (sense and avoid),
    \item ``\emph{an aircraft in the vicinity of an aerodrome must make all turns to the left unless told otherwise}'' (navigation in airspace), and
    \item ``\emph{an aircraft shall not taxi on the apron or the manoeuvering area of an aerodrome without permission}'' (ATC clearance).
\end{itemize}

The formalisation of these statements was performed using temporal logic and an automatic verification tool  known as a \emph{model checker} was employed for verification against a formal model. Naturally, this approach relies critically on (i) the formalisation being correct (i.e. the semantics of the formal specification being consistent with the intended meaning of the natural language expressions) and (ii) on the quality of the model as a genuine abstraction of the real system; if the model does not faithfully represent the actual UAV, no meaningful equivalence can been established.

It can, in practice, be difficult or impossible to formalise some other requirements emanating from regulatory bodies that make reference to human operators (such as pilots) of a certain level of skill being able to perform certain functions. Specifications of this kind are not presently amenable to formalisation.

The requirements considered in the work of Webster et al. are quite specific, but the sense and avoid requirement is standard for aerial vehicles; though its precise phrasing can vary (e.g ``\emph{vigilance shall be maintained by each person operating an aircraft so as to see and avoid other aircraft}'' in \cite{FAA}), the safety specification it is designed to ensure is that of collision freedom and can be readily formalised. 

The collision avoidance requirement is much more safety-critical in manned aircraft than UAVs, which has motivated work on formal verification of collision avoidance for systems equipped with ACAS  (Airborne Collision Avoidance System) X system by J-B Jeannin and co-authors in~\cite{jeannin2015formally}.

\begin{remark}
Safety standards in aviation are often stated in probabilistic terms, e.g.
``\emph{The acceptable level of risk (ALR) for UAS should be consistent
across all types of operations being performed, and no more restrictive than the accepted fatality rates of general aviation}''~\cite[AG 2.1]{FAA}, where fatality rates are measured as the number of deaths per 100,000 flight hours.

Also, it needs to be stressed that the operating environment of both manned aircraft and UAVs affects both the air and ground risks~\cite{FAA}; for example, a falling UAV may cause damage as it impacts the ground in a populated area. Mitigating ground risks remains a \emph{safety} requirement, which at an abstract level essentially amounts to the property of collision avoidance between entities.
\end{remark}

\subsection{Safety Specifications}

Formal requirements often combine \emph{liveness} (in computer science parlance a property that ``something good eventually happens'', such as a system attaining a desirable state) and \emph{safety}, which in computer science is informally understood as a property that ``nothing bad happens'' throughout the operation of the system (e.g. the system never transitions into an undesirable state). A typical example would be a specification for an autonomous vehicle which is required to move from position $A$ to position $B$ in its environment, while avoiding any obstacles along the way. Likewise, a swarm of UAVs may be required to travel to a different location while avoiding collisions among the UAVs along the way.
These common kinds of requirements are sometimes termed \emph{reach-avoid} and may be stated formally with reference to a particular mathematical model. 

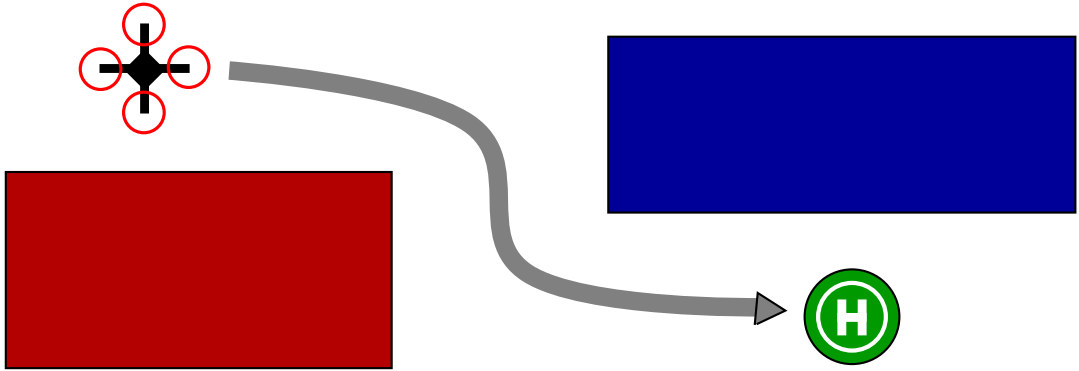
\begin{figure}[h!]
\centering
    \begin{tikzpicture}[thick,scale=0.18, every node/.style={scale=0.18}]

\draw[fill=black]  (-2.5,1.4) rectangle (4,0.9);
\draw[fill=black]  (0.5,4.4) rectangle (1,-2.1);

\draw[very thick, red]  (0.7,-2.1) ellipse (1.5 and 1.5);
\draw[very thick, red]  (-2.5,1.1) ellipse (1.5 and 1.5);
\draw[very thick, red]  (4,1.25) ellipse (1.5 and 1.5);
\draw[very thick, red]  (0.7,4.4) ellipse (1.5 and 1.5);
\draw[fill=black, rotate=45]  (0.2788,1.208) rectangle (2.2788,-0.792);

\draw[fill=red!70!black]  (-9.5,-6.5) rectangle (19,-21);
\draw[fill=green!60!black]  (53,-17.2) ellipse (3.5 and 3.5);
\draw[white, ultra thick]  (53,-17.2) ellipse (2.5 and 2.5);
\draw[fill=white, white]  (52,-16) rectangle (52.5,-18.5);
\draw[fill=white, white]  (53.5,-16) rectangle (54,-18.5);
\draw[fill=white, white]  (52,-17) rectangle (54,-17.5);

\draw[fill=blue!60!black]  (35,3.5) rectangle (69.5,-9.5);

\draw[line width=7pt, gray]  plot[smooth, tension=.7] coordinates {(7,1)    (24.6664,-2.8886) (29.0556,-14) (46,-16.5)};

\draw[fill=gray, rotate=25] (34,-35.5) node (v1) {} -- (35.2,-33.45) -- (36.5,-35.5) -- (v1);
\end{tikzpicture}
    \caption{Satisfying a reach-avoid requirement in an aerial setting}
    \label{fig:reachavoid}
\end{figure}
%

It needs to be stressed that in industrial practice \emph{safety specifications} do not always align with the notion of safety used in computer science. Specifications of safety-critical mission requirements are not limited to properties such as avoiding obstacles or maintaining velocity (or temperature, pressure, etc.) within prescribed operating bounds (i.e. the standard, but narrow view of safety,  as this term is employed in computer science).
Safety-critical requirements can in fact include many kinds of temporal properties, such as:
\begin{enumerate}
    \item \emph{eventuality} -- i.e. a form of liveness requiring that a \emph{target} set of states is eventually attained by the system,
    \item \emph{stability} (of various kinds),
    \item operating within strict \emph{deadlines}, 
    \item 
    \emph{absence of oscillations}.
\end{enumerate}
One may find an example of the latter safety-critical requirement for \emph{avoiding stick-slip oscillations} in a model of a conventional oil well drill string (described in more detail in~\cite{JAR18}). Intuitively, the safety specification for this system could be informally stated as: ``\emph{the drill pipes should not become damaged by the drilling bit repeatedly becoming stuck and unstuck during the drill's operation}''. 
%


Specifications of system behaviour can be made precise using formal logics. Temporal logics in particular have proved very useful as they permit one to express many interesting properties that relate the states of the system to time. Established temporal logics such as Linear Temporal Logic (LTL) were first introduced to specify (and verify) the behaviour of computer programs, but have since come to be used in fields such as robotics to specify motion plans (a good overview of this application of LTL may be found in the article by Plaku and Karaman~\cite{plaku2016motion}). Logics like LTL can express properties of systems in which both state and time are discrete; as such, they can only be applied to verify discrete abstractions of physical systems. 

\begin{figure}[h!]
    \centering
    \begin{tikzpicture}

\draw[fill=blue]  (-3.5,4) rectangle (-3,3.5) node (v1) {};
\draw  (-3,4) rectangle (-2.5,3.5) node (v2) {};
\draw  (-2.5,4) rectangle (-2,3.5) node (v3) {};
\draw  (-2,4) rectangle (-1.5,3.5) node (v4) {};
\draw[fill=blue]  (-1.5,4) rectangle (-1,3.5) node (v5) {};
\draw  (-1,4) rectangle (-0.5,3.5) node (v6) {};
\draw  (-0.5,4) rectangle (0,3.5) node (v7) {};
\draw  (0,4) rectangle (0.5,3.5);

\draw  (-3.5,3.5) rectangle (-3,3);
\draw  (v1) rectangle (-2.5,3);
\draw  (v2) rectangle (-2,3);
\draw  (v3) rectangle (-1.5,3);
\draw  (v4) rectangle (-1,3);
\draw  (v5) rectangle (-0.5,3);
\draw  (v6) rectangle (0,3);
\draw[fill=blue]  (v7) rectangle (0.5,3);

\draw  (-3.5,3) rectangle (-3,2.5);
\draw  (-3,3) rectangle (-2.5,2.5);
\draw  (-2.5,3) rectangle (-2,2.5);
\draw  (-2,3) rectangle (-1.5,2.5);
\draw  (-1.5,3) rectangle (-1,2.5);
\draw  (-1,3) rectangle (-0.5,2.5);
\draw  (-0.5,3) rectangle (0,2.5);
\draw  (0,3) rectangle (0.5,2.5);

\draw[fill=blue]  (-3.5,2.5) rectangle (-3,2);
\draw  (-3,2.5) rectangle (-2.5,2);
\draw  (-2.5,2.5) rectangle (-2,2);
\draw  (-2,2.5) rectangle (-1.5,2);
\draw  (-1.5,2.5) rectangle (-1,2);
\draw  (-1,2.5) rectangle (-0.5,2);
\draw  (-0.5,2.5) rectangle (0,2);
\draw  (0,2.5) rectangle (0.5,2);


\draw  (-3.5,2) rectangle (-3,1.5);
\draw  (-3,2) rectangle (-2.5,1.5);
\draw  (-2.5,2) rectangle (-2,1.5);
\draw  (-2,2) rectangle (-1.5,1.5);
\draw  (-1.5,2) rectangle (-1,1.5);
\draw  (-1,2) rectangle (-0.5,1.5);
\draw  (-0.5,2) rectangle (0,1.5);
\draw[fill=blue]  (0,2) rectangle (0.5,1.5);


\draw  (-3.5,1.5) rectangle (-3,1);
\draw  (-3,1.5) rectangle (-2.5,1);
\draw  (-2.5,1.5) rectangle (-2,1);
\draw[fill=green]  (-2,1.5) rectangle (-1.5,1);
\draw  (-1.5,1.5) rectangle (-1,1);
\draw  (-1,1.5) rectangle (-0.5,1);
\draw  (-0.5,1.5) rectangle (0,1);
\draw  (0,1.5) rectangle (0.5,1);


\node at (-1.75,1.25) {A};
\node at (-3.25,3.75) {B};
\node at (-3.25,2.25) {C};
\node at (-1.25,3.75) {D};
\node at (0.25,3.25) {E};
\node at (0.25,1.75) {F};
\end{tikzpicture}
    \caption{Visiting locations in an environment}
    \label{fig:surveyMission}
\end{figure}
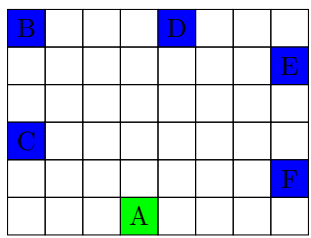
For example, consider an environment illustrated in Fig~\ref{fig:surveyMission} which can represent an abstraction of a floor of a building where the regions near windows are represented by cells coloured in blue and the cell coloured in green represents the region containing a robot which can move in the plane by transitioning into any neighbouring cell in a single time step.
The language of linear temporal logic provides \emph{temporal operators} (typically these are $\Box$ : ``always''; $\Diamond$ : ``eventually''; $\mathcal{U}$ : ``until'', and $X$ : ``next'', i.e. ``in the next state'') which allow us to formally express (i.e. specify) properties of the desired motion of the robot.  For example, the LTL formula $\Diamond \phi$ expresses the property of eventuality of $\phi$, i.e. the fact that \emph{eventually} the system enters a state where formula $\phi$ is true. With this, one can formalise a simple surveillance task in which the robot is required to visit all the regions near the windows on the floor as:
\begin{align*}
   \Diamond\,\mathsf{robAtB} \land \Diamond\,\mathsf{robAtC} \land \Diamond\,\mathsf{robAtD}
   \land \Diamond\,\mathsf{robAtE}
   \land \Diamond\,\mathsf{robAtF}\,,
\end{align*}
where $\mathsf{robAtB}$ represents the proposition ``the robot is at location B'', respectively for locations C, D, E and F.

\begin{remark}
\label{rem:response}
A very common kind of specification follows what is known as a \emph{response pattern}, which specifies that it is always the case that an event P is eventually followed by event Q, which takes the form
\[
\Box \left( P \to \Diamond Q \right)\,.
\]
\end{remark}

While Linear Temporal Logic can be used to formally express many important requirements that arise in practice, it does have limitations. For instance, it is not possible to express the requirement that ``\emph{it is always possible for the robot to reach $D$}'' (using the example in Fig.~\ref{fig:chessboard}) in LTL. Instead, such a requirement can be formally stated using Computational Tree Logic (CTL), which likewise has been applied for specifying autonomous systems~\cite{sirigineedi2011kripke} and features so-called \emph{path quantifiers} $A$ (which may be informally read as ``along all computation paths'') and $E$ (``along at least one computation path'') that are paired with path-specific quantifiers $G$ (``globally along the path'') and $F$ (``eventually along the path''). The above requirement (which corresponds to a so-called \emph{reset property}) can be formalised in CTL as $\mathrm{AG}\, \mathrm{EF}~\mathsf{robAtD}$. Like LTL, CTL can also be used to express response pattern in Rem.~\ref{rem:response} as $\mathrm{AG}(P \to \mathrm{AF}\,Q)$; however, the property expressed by the LTL formula $\Diamond\,\Box\, \mathsf{robAtD}$ cannot be expressed in CTL. A logic that known as CTL$^*$ serves to combine the expressive power of LTL and CTL.

The interested reader may find an accessible introduction to  the temporal logics LTL, CTL and CTL$^*$ in the excellent book by Huth and Ryan~\cite{huthRyan}. The merits and important differences between the logics LTL and CTL, along with their practical implications are also helpfully described by Vardi in~\cite{VardiTACAS01}.

In many practical applications (such as in real time systems and cyber-physical systems), specifications involve  \emph{real} time which does not advance in discrete steps, as well as other continuously evolving variables. This motivated the development of logics such as \emph{Metric Temporal Logic} (MTL)~\cite{DBLP:journals/rts/Koymans90} and \emph{Signal Temporal Logic} (STL)~\cite{DBLP:conf/formats/MalerN04}, among others, in which the temporal operators feature time constraints. 
For instance, a specification of a collision avoidance property between a vehicle and an obstacle expressing that the vehicle must maintain some minimum safe distance $d_{\min}>0$ from the obstacle at all times in the future can be written down formally as:
\[
\Box_{[0,\infty)}~\mathsf{dist}(vehicle, obstacle)\geq d_{\min}\,,
\]
where the temporal operator $\Box_{[0,\infty)}$ in the above formula expresses \emph{necessity} and the formula may be read as ``at all times in the future $\mathsf{dist}(vehicle,obstacle)\geq d_{\min}$ holds''. The temporal operator~$\Diamond$, which expresses \emph{eventuality}, can feature a finite timing constraint allowing us to write 
\[
\Diamond_{[0,T]}~\mathsf{Target}\,,
\]
which informally says that the set of states described by the formula $\mathsf{Target}$ is ``eventually attained within $T$ time units''. A reach-avoid specification can be expressed using a conjunction, i.e.
\[
\left(\Diamond_{[0,T]}~\mathsf{Target}\right)~\land~\left(\Box_{[0,T]}~\mathsf{Safe}\right)\,,
\]
where \textsf{Safe} is a formula describing the set of safe states of the system.
\begin{remark}
We should note that the common \emph{reach-avoid} specifications can be stated in purely set theoretic terms, i.e. by explicitly providing a set \emph{Reach} of states that the system must enter (i.e. states satisfying \textsf{Target})  and a set \emph{Avoid} of states that the system must avoid on its way to a state inside \emph{Reach} (i.e. states satisfying \textsf{Safe}).
\end{remark}

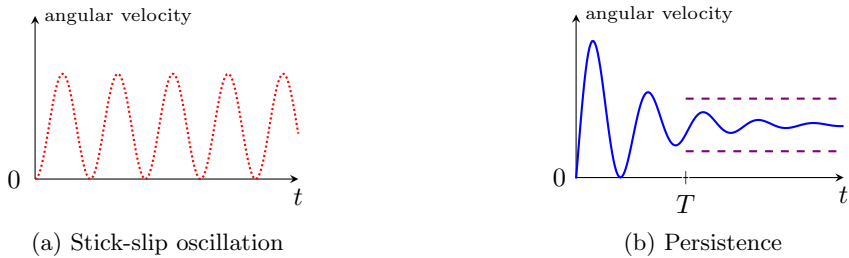
\begin{figure}[h!]
    \begin{subfigure}[t]{0.45\textwidth}
        \centering
    \begin{tikzpicture}[domain=1.57:5+1.57, scale=0.7] 
    \draw[->, >=stealth] (-0.0+1.57,0) -- (5+1.57,0) node[below] {$t$}; 
	\node[draw=white] at (-0.4+1.57,0) {$0$};
    \draw[->,>=stealth] (-0.0+1.57,-0.0) -- (-0.0+1.57,3.1) node[right] {\scriptsize angular velocity};
    \draw[color=red, samples=200, thick, densely dotted]   plot (\x,{1*cos(6.0*\x r)+1})    node[right] {}; 
  \end{tikzpicture}
        \caption{Stick-slip oscillation\label{subfig:stickslip}}
    \end{subfigure}%
    ~ 
    \begin{subfigure}[t]{0.45\textwidth}
        \centering
    \begin{tikzpicture}[domain=-0.082:5, scale=0.7] 
    \draw[->, >=stealth] (-0.1,0) -- (5,0) node[below] {$t$}; 
	\node[draw=white] at (-0.4,0) {$0$};
	\node[draw=white, inner sep=0pt] at (2,-0) {\tiny $|$};
	\node[draw=white, inner sep=0pt] at (2,-0.5) {$T$};
    \draw[->,>=stealth] (-0.082,-0.0) -- (-0.082,3.1) node[right] {\scriptsize angular velocity};
    \draw[color=blue, samples=200, thick]   plot (\x,{2*exp(-0.9*\x)*sin(6.0*\x r)+1})    node[right] {}; 
    \draw[color=violet, samples=200, thick, dashed, domain=2:5]   plot (\x,{1.5})    node[right] {}; 
    \draw[color=violet, samples=200, thick, dashed, domain=2:5]   plot (\x,{0.5})    node[right] {}; 
  \end{tikzpicture}
        \caption{Persistence\label{subfig:persistence}}
    \end{subfigure}
   \caption{Absence of stick-slip oscillations~\label{fig:stickslip}}
\end{figure}

Properties such as absence of stick-slip oscillations in an oil drill correspond to a temporal property known as \emph{persistence} (illustrated in Fig.~\ref{subfig:persistence}), which can be expressed using a nested combination of temporal operators: 
\[
\Diamond_{[0,T]}~\Box_{[0,\infty)}~\mathsf{Steady}\,,
\]
informally read as ``\emph{eventually, within $T$ time units, the system enters the $\mathsf{Steady}$ region and remains within it}'', where the formula \textsf{Steady} represents bounds on the angular velocity with the lower bound being a positive so as to ensure the angular velocity does not drop to zero which could cause stick-slip oscillations (as in Fig.~\ref{subfig:stickslip}).

Temporal logics such as LTL and STL have been applied in robotics to specify and solve problems in motion  control~\cite{chen2012ltl}, as well as multi-agent motion planning~\cite{sun2022multi}. Other formal logics, such as ATL (Alternating-time Temporal Logic)~\cite{alur2002alternating}, permit one to formalise specifications that make reference to cooperation among entities, such as e.g. ``agents 1 and 2 can cooperate to ensure that the system never enters a fail state''~\cite{van2012logics}.

Differential dynamic logic (\textsf{dL}) was first introduced by Platzer~\cite{platzer2008differential} and provides a unified specification and verification framework for hybrid systems. In this framework, a safety specification can take the form
\[
\mathrm{Init} \to [ \alpha ]~\mathrm{Safe}\,,
\]
where $\mathrm{Init}$ is a formula representing the set of initial states, $\mathrm{Safe}$ is a formula representing the set of safe states of the system, and $\alpha$ is a \emph{hybrid program} which provides an operational model of the system itself; in this formula, the modal operator $[\alpha]~\mathrm{Safe}$ means that the formula $\mathrm{Safe}$ holds after all executions of hybrid program $\alpha$. Differential \emph{temporal} dynamic logics have also been developed~\cite{dtl,dtl2}, which extend \textsf{dL} with temporal modalities $\Box$ and $\Diamond$.
\begin{remark}
    Having an operational model of the system be part of the formula that provides the specification in dynamic logic is rather different to the situation with temporal logics such as LTL, where one typically has a finite state model of the system $S$, separately an LTL property $P$ which provides the specification, and wishes to check whether the system satisfies the specification (often written $S \models P$).
\end{remark}

As a modelling formalism, hybrid programs provide constructs for assigning values to variables, creating loops, conditionals, as well as letting variables evolve continuously according to \emph{differential equations}. 
    Hybrid programs thus provide an alternative to \emph{hybrid automata}~\cite{henzinger2000theory}, which is another popular model for cyber-physical systems.

Differential dynamic logic has been used in numerous works to specify (and formally verify) the behaviour of terrestrial vehicles~\cite{mitsch2013provably,bohrer2019formal};  Selvaraj et al. recently applied this formal framework to safe autonomous driving~\cite{selvaraj2022formal} and describe some important nuances in formal modelling~\cite{selvaraj2022not}.

It is interesting to note that Signal Temporal Logic was originally introduced for \emph{monitoring} properties of signals, rather than for verification of these properties; it has proved convenient and has seen some use in industry. In contrast, differential dynamic logic \textsf{dL} was from the outset designed to support a calculus for constructing formal proofs. Recent work by Ahmad and Jeannin~\cite{ahmad2021program} introduced a Signal Temporal Dynamic Logic (ST\textsf{dL}) which attempts to bridge these differences.

\subsection{Hyperproperties}
Not all specifications can be directly stated as properties using temporal logics such as LTL or STL. The notion of  \emph{hyperproperties} was first introduced by Clarkson and Schneider~\cite{clarkson2010hyperproperties} in their study of security properties and intuitively requires more than one trace (i.e. execution of the system) in order to check, in contrast to standard trace properties that one can express in logics like LTL.

\begin{remark}
In formal methods, the notion of a \emph{property} is usually identified with a \emph{set of traces} (a subset of all possible traces $Tr$). Thus, to say that a system $S$ satisfies property $P$ (i.e. $S \models P$) simply means $S \subseteq P$. The computer science notion of safety is an example of such a trace property. In contrast, \emph{hyperproperties} are subsets of the powerset $2^{Tr}$.
\end{remark}

\begin{figure}[h!]
    \centering
    \begin{tikzpicture}

\draw[fill=red]  (-3.5,4) rectangle (-3,3.5) node (v1) {};
\draw  (-3,4) rectangle (-2.5,3.5) node (v2) {};
\draw  (-2.5,4) rectangle (-2,3.5) node (v3) {};
\draw  (-2,4) rectangle (-1.5,3.5) node (v4) {};
\draw  (-1.5,4) rectangle (-1,3.5) node (v5) {};
\draw  (-1,4) rectangle (-0.5,3.5) node (v6) {};
\draw  (-0.5,4) rectangle (0,3.5) node (v7) {};
\draw  (0,4) rectangle (0.5,3.5);

\draw  (-3.5,3.5) rectangle (-3,3);
\draw  (v1) rectangle (-2.5,3);
\draw  (v2) rectangle (-2,3);
\draw  (v3) rectangle (-1.5,3);
\draw  (v4) rectangle (-1,3);
\draw  (v5) rectangle (-0.5,3);
\draw  (v6) rectangle (0,3);
\draw  (v7) rectangle (0.5,3);

\draw[fill=black]  (-3.5,3) rectangle (-3,2.5);
\draw[fill=black]  (-3,3) rectangle (-2.5,2.5);
\draw[fill=black]  (-2.5,3) rectangle (-2,2.5);
\draw  (-2,3) rectangle (-1.5,2.5);
\draw  (-1.5,3) rectangle (-1,2.5);
\draw  (-1,3) rectangle (-0.5,2.5);
\draw  (-0.5,3) rectangle (0,2.5);
\draw  (0,3) rectangle (0.5,2.5);

\draw[fill=black]  (-3.5,2.5) rectangle (-3,2);
\draw[fill=black]  (-3,2.5) rectangle (-2.5,2);
\draw  (-2.5,2.5) rectangle (-2,2);
\draw  (-2,2.5) rectangle (-1.5,2);
\draw  (-1.5,2.5) rectangle (-1,2);
\draw  (-1,2.5) rectangle (-0.5,2);
\draw  (-0.5,2.5) rectangle (0,2);
\draw  (0,2.5) rectangle (0.5,2);


\draw  (-3.5,2) rectangle (-3,1.5);
\draw  (-3,2) rectangle (-2.5,1.5);
\draw  (-2.5,2) rectangle (-2,1.5);
\draw  (-2,2) rectangle (-1.5,1.5);
\draw  (-1.5,2) rectangle (-1,1.5);
\draw  (-1,2) rectangle (-0.5,1.5);
\draw  (-0.5,2) rectangle (0,1.5);
\draw  (0,2) rectangle (0.5,1.5);


\draw  (-3.5,1.5) rectangle (-3,1);
\draw  (-3,1.5) rectangle (-2.5,1);
\draw  (-2.5,1.5) rectangle (-2,1);
\draw  (-2,1.5) rectangle (-1.5,1);
\draw  (-1.5,1.5) rectangle (-1,1);
\draw[fill=black]  (-1,1.5) rectangle (-0.5,1);
\draw[fill=black]  (-0.5,1.5) rectangle (0,1);
\draw[fill=black]  (0,1.5) rectangle (0.5,1);


\draw  (-3.5,1) rectangle (-3,0.5);
\draw  (-3,1) rectangle (-2.5,0.5);
\draw  (-2.5,1) rectangle (-2,0.5);
\draw  (-2,1) rectangle (-1.5,0.5);
\draw  (-1.5,1) rectangle (-1,0.5);
\draw  (-1,1) rectangle (-0.5,0.5);
\draw  (-0.5,1) rectangle (0,0.5);
\draw[fill=black]  (0,1) rectangle (0.5,0.5);


\draw  (-3.5,0.5) rectangle (-3,0);
\draw  (-3,0.5) rectangle (-2.5,0);
\draw  (-2.5,0.5) rectangle (-2,0);
\draw  (-2,0.5) rectangle (-1.5,0);
\draw  (-1.5,0.5) rectangle (-1,0);
\draw[fill=green]  (-1,0.5) rectangle (-0.5,0);
\draw  (-0.5,0.5) rectangle (0,0);
\draw  (0,0.5) rectangle (0.5,0);


\node at (-0.75,0.25) {$s_6$};
\node at (-3.25,3.75) {$s_{57}$};

\end{tikzpicture}
    \caption{Taking an optimally short path as part of the specification}
    \label{fig:chessboard}
\end{figure}

To appreciate the difference between hyperproperties and trace properties it is helpful to consider an example. Suppose we have a system which provides an abstraction of a terrestrial robot navigating an environment illustrated in Fig.~\ref{fig:chessboard}, where rectangular cells represent the regions where the robot can be situated and correspond discrete states of the system. The system is initialised in state $s_6$ (i.e. the robot starts operating from the region in the green cell in Fig.~\ref{fig:chessboard}) and is required to transition to state $s_{57}$ (i.e. the region represented by the red cell) \emph{in the fewest possible number of steps} without entering any of the forbidden regions coloured in black; the system may transition into any of its neighbouring states (cells) in a single step. In this example, it is possible to determine from a single trace (execution, i.e. a sequence of states such as $s_6s_5s_4s_{12}\dots$) of the system whether the required state $s_{57}$ is attained at some point, but it is impossible to judge whether this trace is optimally short; in order to do this, one requires access to all the other traces of the system which attain $s_{57}$ in order to compare them. Optimality therefore provides an example of a hyperproperty.
This and other applications of hyperproperties in robotics, in particular in robot motion planning are described in the recent work by Wang et al.~\cite{wang2020hyperproperties}.

Hyperproperties are of interest in security because they enable one to specify e.g. the susceptibility of a system to side channel attacks~\cite{nguyen2017hyperproperties}.  Logics for hyperproperties have been developed by Clarkson et al.~\cite{clarkson2014temporal}. In cyber-physical systems Nguyen et al.~\cite{nguyen2017hyperproperties} explored the use of hyperproperties for specifying security and stability requirements and proposed a new formal logic  HyperSTL, which extends STL.


\subsection{Issues with Temporal Logic Formalisations and Tool Support}
It has been appreciated for some time that -- even when there is no ambiguity -- fairly innocent informal specifications of system behaviour can require rather unwieldy temporal logic formulas. Dwyer et al.~\cite{dwyer1999patterns} give the following example of a lift behaviour specification: ``\emph{Between the time an elevator is called at a floor
and the time it opens its doors at that floor, the elevator can arrive at that floor at most twice}.'' Infamously, formalising this specification in LTL (using the ``until'' temporal operator $\mathcal{U}$, in addition to $\Box$ and $\Diamond$) one arrives at:
\begin{align*}
    & \square~( ( call \land \lozenge open) \to \\
    & \quad \quad ( ( \lnot atfloor \land \lnot open)~\mathcal{U} \\
    & \quad \quad \quad (open \lor ((atfloor \land \lnot open)~\mathcal{U} \\
    & \quad \quad \quad \quad (open \lor ((\lnot atfloor \land  \lnot open)~\mathcal{U}\\
    & \quad \quad \quad \quad \quad (open \lor ((atfloor \land \lnot open)~\mathcal{U} \\
    & \quad \quad \quad \quad \quad \quad (open \lor (\lnot atfloor~\mathcal{U}~open)))))))))\,.
\end{align*}
The work of Dwyer et al. in~\cite{dwyer1999patterns} sought to develop a \emph{specification pattern system} to organise commonly occurring specification \emph{patterns} into a hierarchy so as to facilitate formalisation of common high-level requirements.

Tools such as \emph{FRET} (Formal Requirements Elicitation Tool)~\cite{giannakopoulou2020formal} have been developed at NASA to help with the formalisation of requirements. The idea behind this approach is to adopt a restricted natural language with precise semantics in which users can write informal requirements which can then be formalised in temporal logic.
More recently, He et al.~\cite{dejanSTL} developed a tool DeepSTL which is designed to translate informal English language requirements into Signal Temporal Logic specifications. 
 
\section{Mission Specifications}
As remarked in Sec.~\ref{sec:regulationSpecs}, regulations provide a large corpus of \emph{de facto} specifications for autonomous systems; these can be thought of as constraints on their desired behaviour. If it is possible to respect these constraints, the main task lies in  specifying the \emph{mission} of an autonomous system.
Mission specifications are generally not restricted to autonomous systems (but certain missions may e.g. be deemed too dangerous for a human operator to be involved, or take place in an inhospitable environment which may require full autonomy) and span a very wide spectrum of applications. Below we list some existing work on formal specification of missions.

\paragraph{Catalogues of robotic missions}
Menghi et al. created a catalogue of mission specifications that are commonly employed in robotics~\cite{menghi2019specification} and provide temporal logic \emph{templates} for these missions (the templates are available online \href{https://dsg.tuwien.ac.at/team/ctsigkanos/patterns/}{https://dsg.tuwien.ac.at/team/ctsigkanos/patterns/}). For example, the intent of \emph{fair patrolling} mission is defined by the authors in natural language as ``\emph{This pattern requires a robot to keep visiting a set of locations in a fair way, i.e., the robots patrols the locations by keeping the number of times every area is patrolled equal.}'', which is subsequently formalised in LTL and CTL.
Recent work by Mallozzi et al. on contract-based robotic mission specification~\cite{mallozzi2020crome} likewise employed Linear Temporal Logic in formalising the requirements.

\paragraph{Surveillance}
The work by Smith et al.~\cite{smith2011optimal} considered path planning for surveillance missions and employed Linear Temporal Logic (LTL). An example mission for a robot in a (given) model of a terrestrial environment is~\cite{smith2011optimal}: \emph{``Repeatedly gather data at locations g1, g2, and g3. Upload data at either
u1 or u2 after each data-gather. Follow the road rules, and avoid the road connecting i4 to i2.''} (where the road intersections in the environment are labelled with $i$, the gathering locations are labelled with $g$ and the upload locations are labelled with $u$).
Such a mission can be formally written down using LTL formulas (see~\cite[\S 5]{smith2011optimal}). Furthermore, the authors sought to generate \emph{optimal} robot paths; however, optimality is not directly expressed in the formal logic as part of the specification and is rather ensured by the algorithm developed in this work.


\paragraph{Hitting targets}
In~\cite{karaman2008complex}, Karaman and Frazzoli  consider the
\emph{Multiple-UAV Mission Planning problem}, in which one is given a set of targets and a set of UAVs and one is required to assign the targets to
UAVs in an optimal manner (as well as generate the optimal paths that each UAV has to follow). A case study featuring a mission specification inspired by a military scenario is described in~\cite{karaman2008complex}: ``\emph{The mission in this case is to destroy either T1 and T3,
or T2 and T3. Thus making a way through for the infantry
from T1 side or T2 side. There is also another specification
which is that if the rescued unit escapes from T2 side and
reaches the friendly base C2 then V4 must meet the rescued
unit there with necessary health supplies}'' (here the UAVs are denoted by $V$, the targets by $T$). The overall specification can be formalised as a conjunction of LTL constraints (see~\cite[\S V]{karaman2008complex}). In this work, once again, the LTL specification is viewed as a \emph{constraint}, subject to which an optimisation problem can be encoded and solved, i.e. optimality itself is not expressed within the logic formalism.

\paragraph{Station keeping}
Martin et al.~\cite{martin2017formal} consider \emph{station keeping manoeuvres} of a Dubins vehicle. A station keeping manoeuvre requires that ``\emph{the vehicle reaches a neighbourhood of its station in finite time
and remains in it while waiting for further instructions.}'' Such a manoeuvre is a combination of a liveness (eventuality) property (i.e. the vehicle must reach the neighbourhood of its station in finite time) and a safety property (i.e. the vehicle is to remain in the neighbourhood of its station once it is there). The combination of the two requirements results in a \emph{persistence} property which can be stated and verified in differential dynamic logic, as done in~\cite{martin2017formal} (although this property could also be stated using a logic such as STL).

\subsection{Challenges for Verification}

Formalisms like temporal logics often allow us to precisely write down (i.e. specify) requirements for the behaviour of systems. The task of \emph{verifying} these requirements for a given formal model of a system can often be a far more difficult task, which is generally \emph{undecidable} and thus cannot be performed automatically for many interesting systems. %
Nevertheless, verification of formal requirements is sometimes possible

%

\begin{enumerate}
\item
For purely discrete models of systems, formal specifications in temporal logics such as LTL can be verified \emph{automatically} using  model checking (e.g. see~\cite{huthRyan}).
As a purely automated verification technology, model checking has proved tremendously successful and has been applied in large industrial verification problems (particularly in hardware verification). A number of mature model checking tools are available, such as NuSMV~\cite{cimatti2002nusmv}, nuXmv~\cite{cavada2014nuxmv} and SPIN~\cite{holzmann1997model}.

The challenge in applying model checking as a verification technology to cyber-physical systems lies in the fact that one cannot \emph{in practice} compute reachable sets of continuous systems exactly (even up to finite time bound) and is forced to over-approximate. These and other difficulties are described in the work of Fehnker and Krogh in~\cite{fehnker2004hybrid}. 

\item Verification of \emph{bounded-time} properties, such as safety for a bounded duration of time, eventuality and reach-avoid properties is often possible by employing one of a family of methods -- broadly known as \emph{reachability analysis}. Bounded-time verification is closely related to a verification approach known as \emph{bounded model checking} (BMC)~\cite{biere2009bounded}.  
Some of the more successful recent verification tools for \emph{continuous} and \emph{hybrid} dynamical systems based on reachability analysis are \emph{SpaceEx} by Frehse et al.~\cite{frehse2011spaceex}, which works for linear systems, and \emph{Flow*} by Chen et al.~\cite{chen2013flow} which can also handle non-linear dynamics.

\item An alternative verification technology to model checking is known as \emph{theorem proving}. Verification of  important properties such as maintaining a safe distance from obstacles at all times can be achieved by creating a formal proof of such a property inside a theorem prover (also known as an \emph{interactive proof assistant}). Safety properties are typically require
finding appropriate \emph{invariants}, which in itself can be a difficult task. 

The theorem proving approach to verification is much less automatic than model checking and typically requires a considerable level of expertise to be productive. The amount of effort involved in constructing formal proofs has historically hindered the use of theorem provers in industry. However, it needs to be stressed that theorem provers provide an unrivalled level of assurance because they can produce fully rigorous formal proofs (in contrast to model checking tools).
\end{enumerate}

Some noteworthy efforts in cyber-physical system verification using theorem provers are those pursued in the KeYmaera~X prover~\cite{DBLP:conf/cade/FultonMQVP15} (developed at Carnegie Mellon University in a group led by A. Platzer), and the Hybrid Hoare Logic (HHL) prover~\cite{DBLP:conf/icfem/WangZZ15} (developed at the Chinese Academy of Sciences by N. Zhan, S. Wang, and others).
%
%

Lamport's temporal logic of actions~\cite{lamport1994temporal} was originally developed for reasoning about concurrent systems (but has also been applied to model and reason about physical systems~\cite{lamport1992hybrid}) and enjoys extensive tool support in the form of the TLA+ Toolbox~\cite{kuppe2019tla+}; this framework is notable for its successful application in the computer industry (see e.g.~\cite{newcombe2014amazon}).

\paragraph{Specifications for Neural Networks} The increasing prevalence of components such as neural networks in autonomous system designs presents a huge challenge to formal verification, not least because a specification for the behaviour of these components is typically absent. Indeed, some of the main issues with the use of machine learning components in autonomous systems stem from the difficulty of precisely specifying their intended behaviour. This problem brought on by the \emph{absence of a specification for the components of a system} (e.g. lack of specification for a neural net component) even when a formal specification exists for the overall system (e.g. given by a temporal logic formula) is discussed in the work of Seshia and co-workers~\cite{DBLP:conf/atva/SeshiaDDFGKSVY18}, who argue that the overall system specification should serve as a starting point from which one can try to \emph{derive} component specifications.

\section{Control System Specifications}

\begin{figure}[h]
  \centering
  \includegraphics[width=0.9\textwidth]{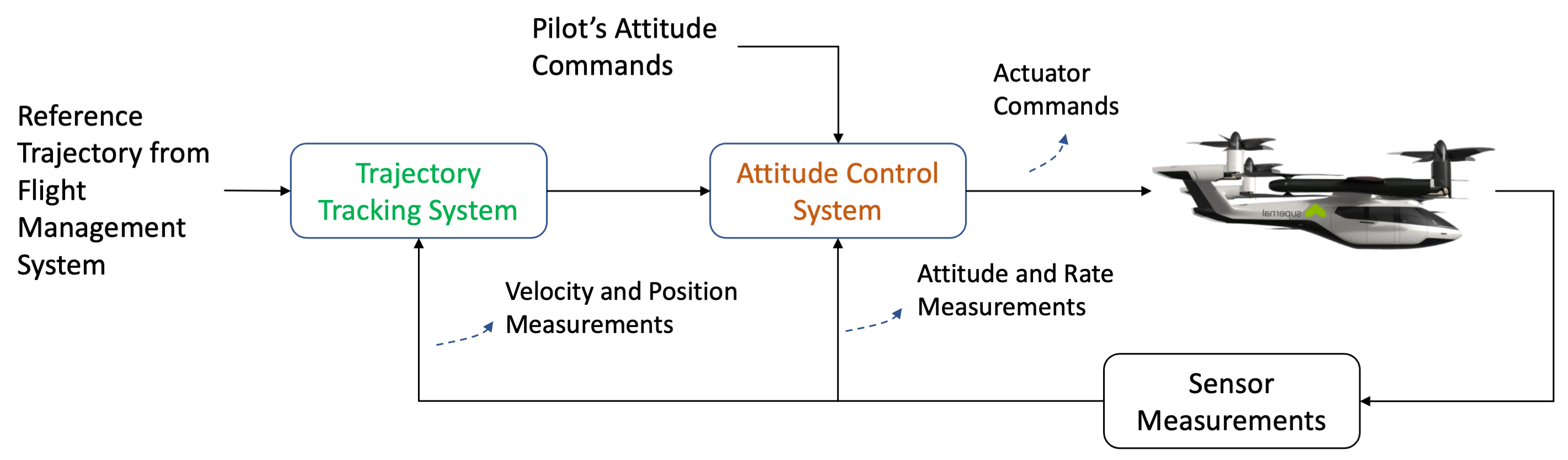}
  \caption{Overview of the cascaded flight control system.}
  \label{fig:fcs_overview}
\end{figure}

In autonomous systems, control algorithms are utilised to provide stability of the whole system and meet the signal tracking requirements to perform specified mission with minimum error. These requirements for the control systems are directly related to operational safety and should be defined clearly for an effective validation and verification process. For an advanced air mobility (AAM) vehicle, schematics of a closed-loop system is given in Figure \ref{fig:fcs_overview}. As shown in this figure, there are two fundamental loops that consist of the flight control system; a) Inner-loop attitude control system, and b) outer-loop trajectory tracking system. 
In the following sections, we briefly review some control specifications for both the inner-loop and outer-loop controllers.

\subsection{Attitude Control System Specifications}
It is generally recognised that the use of rigorous analysis methods at the beginning of requirement development phase of the control system design process significantly reduces the software development costs. In \cite{gross2017formal}, an approach is presented in which common spacecraft system requirements such as actuator limits, steady state error, settling time, rise time and overshoot are formalised. The proposed approach enables analysis of the requirements in early stages of system development. Formalisation and analysis of aerospace control requirements have been studied for decades but the focus has been on the high level requirements and mode logic. In contrast, the focus in~\cite{gross2017formal} is on providing a formal description for low-level control system requirements. 
    
Some hypothetical requirements for an attitude control system with four wheel Reaction Wheel Array (RWA) are described in~\cite{gross2015evaluation, gross2016evaluation, gross2017formal} which we will briefly review below. The first two requirements for constraining the outputs of a control system are designed to  prevent damage to the spacecraft and are given as follows:

    \begin{small}
\begin{itemize}
    \item R01: ``The commanded time rate of change of angular velocity shall not exceed the maximum allowable angular acceleration of the reaction wheel.''
    \item R02: ``The commanded angular velocity shall not exceed the maximum allowable angular velocity of the reaction wheel''
\end{itemize}
\end{small}
    Requirements such as these represent examples of safety specifications where variables are required to never exceed certain given bounds; these bounds in effect represent global invariants and properties such as these can be readily formalised in logics such as STL and \textsf{dL}.
    
Other kinds of requirements considered by Gross in~\cite{gross2017formal}  feature a safety element while also specifying aspects of the expected \emph{performance} of the attitude control system:

\begin{small}
    
\begin{itemize}
    \item R03: ``The pointing accuracy shall be less than 1 degree as a threshold and 0.08 degrees as an objective.''
    \item R04: ``The pointing range about the z axis shall be 0 to 360 degrees.''
    \item R05: ``The pointing range about the y axis shall be 0 to 360 degrees.''
    \item R06: ``The pointing range about the x axis shall be 0 to 360 degrees.''
    \item R07: ``The maximum slew rate shall be $>$ 3 deg/sec as a threshold and $>$ 7 deg/sec as an objective.''
    \item R08: ``After settling, the drift rate shall be $<$ 3 deg/min as a threshold and $<$ 1 deg/min as an objective.''
    \item R09: ``After settling, the total drift shall be $\le$ 0.5 degrees as a threshold and $\le$ 0.1 degrees as an objective.''
    \item R10: ``The 5\% settling time shall be $\le$ 5 minutes as a threshold and $\le$ 2 minutes as an objective.''
    \item R11: ``The 2\% settling time shall be $\le$ 7 minutes as a threshold and $\le$ 3 minutes as an objective.''
    \item R12: ``The rise time shall be $\le$ 5 minutes as a threshold and $\le$ 2 minutes as an objective.''
    \item R13: ``The percent overshoot shall be $\le$ 50\% as a threshold and $\le$ 25\% as an objective.''
\end{itemize}
\end{small}

While ensuring that the thresholds (i.e. bounds) given in these requirements are respected at all times is safety-critical, the ``objective'' part of these requirements describe the desired performance characteristics, which are highly desirable but not safety-critical. This presents a challenge: while it is often possible to express the performance requirements formally, it is important to distinguish between those that are safety-critical versus those that are not. 

An interesting potential solution to this issue involves creating a \emph{hierarchy} of specifications, as described in the recent work of Berducci et al.~\cite{berducci2021hierarchical}, who treat a task specification as an \emph{ordered set} of safety, target and comfort requirements, where the order reflects the level of criticality. The fact that comfort (or performance) requirements may be violated can be addressed using so-called \emph{quantitative semantics} of logics such as STL (e.g. see~\cite{donze2013efficient}).
Quantitative semantics of STL has also been recently employed by Chen et al.~\cite{chen2022stl}) to formalise and reason about the property of \emph{resiliency} in cyber-physical systems.

The practical use of formalisms such as temporal logic is as yet largely unfamiliar outside of computer science and its use in domains such as aerospace is presently very limited. Furthermore, instead of formal verification, extensive \emph{simulation}/testing in commonly employed in the certification process in practice. To give a sense of perspective on the current state of specifications employed in practice, the following section will briefly describe specifications based on so-called Mission-Task Elements.


\subsection{Trajectory Tracking System Specifications: A Mission-Task Element Approach}
For rotary-wing aerial vehicles, flight test manoeuvres are provided as \emph{Mission-Task Elements} (MTEs) in ADS-33E-PRF. The overall assessment of the rotorcraft's ability to perform certain tasks that constitute its mission, such as hover, landing, acceleration-deceleration, is performed by utilising MTEs and this process results in assigning a level of \emph{Handling Qualities} (HQ). A list of applicable MTEs for full-scale rotorcrafts is given in the table below.


\begin{table}[h!]
\centering
\begin{tabular}{@{}cc@{}}
\toprule
\multicolumn{2}{c}{MTEs for Full-scale Rotorcrafts}                 \\ \midrule
\multicolumn{1}{c|}{Hover}              & Acceleration/Deceleration \\
\multicolumn{1}{c|}{Landing}            & Sidestep                  \\
\multicolumn{1}{c|}{Hovering Turn}      & Transient Turn            \\
\multicolumn{1}{c|}{Pirouette}          & Pullup/Pushover           \\
\multicolumn{1}{c|}{Depart/Abort}       & Turn to target            \\
\multicolumn{1}{c|}{Lateral Reposition} & Decelerating Approach     \\
\multicolumn{1}{c|}{Slalom}             & Missed Approach           \\ \bottomrule
\end{tabular}
\end{table}

As a descriptive example for full-scale rotorcraft MTEs, we take a brief look at the performance standards of hover flight below.

\subsubsection*{Performance Standards for Hover Flight Phase}

\begin{figure}[h]
  \centering
  \includegraphics[width=0.7\textwidth]{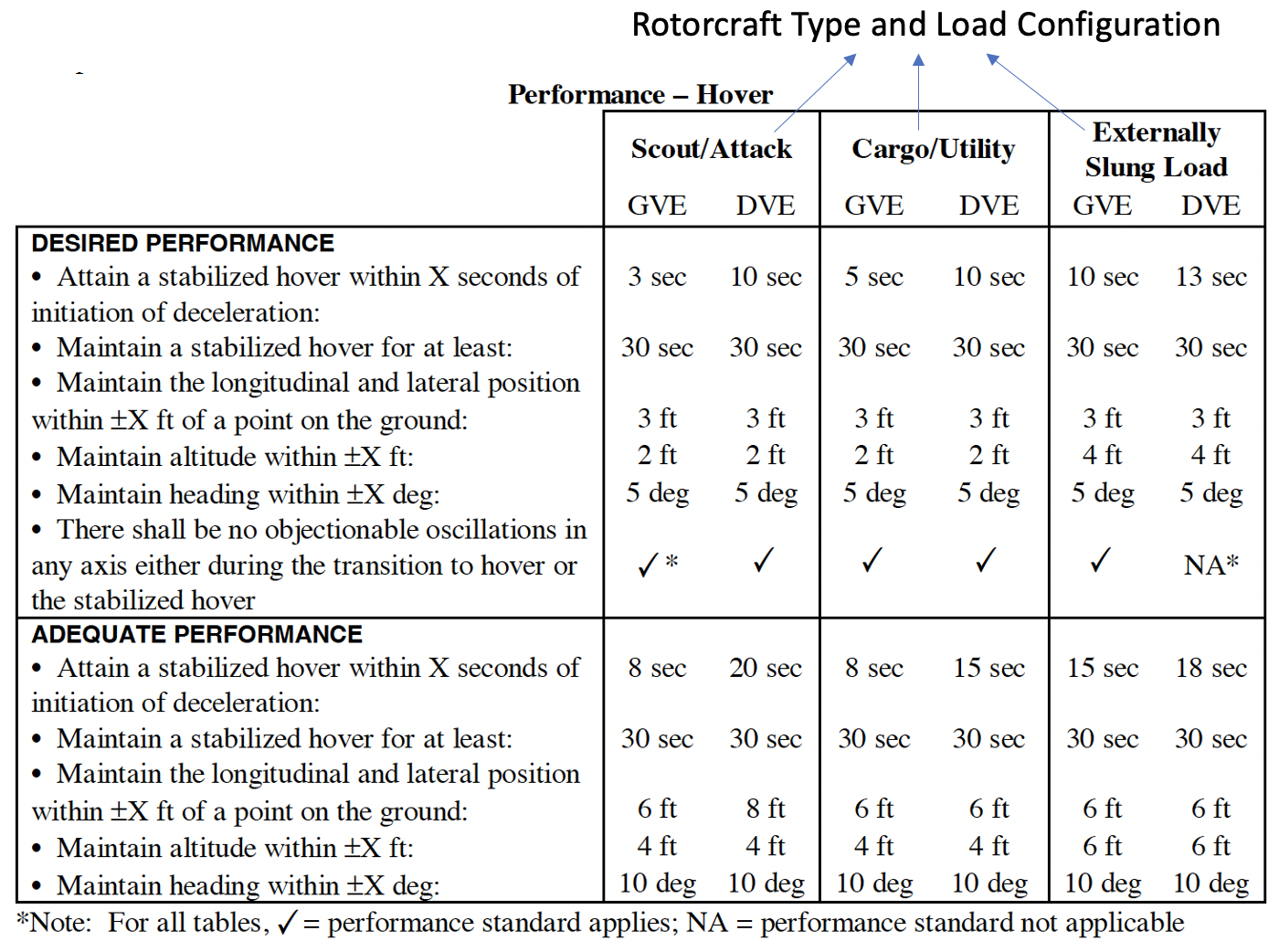}
  \caption{Performance Specifications for Hover Flight.}
  \label{fig:hoverSpecs}
\end{figure}

\begin{itemize}
\item \textit{Objectives}: Checking the ability of transition from forward flight to hover flight with precision and adequate aggressiveness level.

\item \textit{Performance Standards}: Accomplishing the transition to hover flight while meeting the performance requirements given in Figure \ref{fig:hoverSpecs}.
\end{itemize}

\subsubsection*{Scaled-down MTEs for Autonomous Aerial Vehicles}

Although the MTEs in ADS-33E-PRF are designed for piloted rotorcrafts, they have also been utilised to evaluate the trajectory tracking capability of the UAVs \cite{yuksek2020system, ivler2018control}. Here, kinematic scaling, which is a function of maximum cruise velocity of the rotorcraft, is performed to scale-down the original MTEs to obtain applicable missions for UAVs. Here, kinematic scaling is applied to calculate spatial, velocity and time scale factors.

It is also required to define an \emph{objective function} to score the MTE completion performance of the UAV. A weighted sum $(L)$ of the aggressiveness, trajectory tracking performance, robustness is defined as given in Eq. (\ref{eq:objectiveFcn}) to assess the manoeuvre. Then, $L$ is used to calculate the Trajectory Tracking and Aggressiveness (TTA) score $\phi_{TTA}$ as given in Eq.(\ref{eq:tta_score}). Here, $\alpha$ is aggressiveness level of the maneuver, $(\epsilon)$ is trajectory tracking error, and $R$ is robustness. TTA score is between 0 and 100.

\begin{equation} \label{eq:objectiveFcn}
    L = w_{\alpha} \frac{\alpha - \alpha_G}{\alpha_B-\alpha_G} + w_{\epsilon} \frac{\epsilon - \epsilon_G}{\epsilon_B - \epsilon_G} + w_R \frac{R-R_G}{R_B-R_G}
\end{equation}

\begin{equation}\label{eq:tta_score}
    \phi_{TTA}(L) = \frac{200}{1+e^L}
\end{equation}

The quantities $w_a, w_{\epsilon}, w_R$ are weights used to define the relative importance of individual objectives. Subscripts `$B$' and `$G$' stand for best and good possible values. 

After calculating the TTA score by using above mentioned equations, the overall system performance for specific manoeuvres is evaluated based on the score tables given in Table \ref{tab:perfLevels_nominal} and \ref{tab:perfLevels_aggressive} for nominal and aggressive flights, respectively \cite{ivler2018control}.

\begin{table}[]
\centering
\caption{Performance levels for nominal manoeuvring mission}
\begin{tabular}{|c|c|c|}
\hline
Maneuver                                                      & \begin{tabular}[c]{@{}c@{}}Desired TTA Score \\ (Level 1)\end{tabular} & \begin{tabular}[c]{@{}c@{}}Adequate TTA Score \\ (Level 2)\end{tabular} \\ \hline
\begin{tabular}[c]{@{}c@{}}Lateral Reposition\end{tabular} & $\phi_{TTA} \ge 75$                                            & $75 \ge \phi_{TTA} \ge 70$                          \\ \hline
Depart Abort                                                  & $\phi_{TTA} \ge 75$                                            & $75 \ge \phi_{TTA} \ge 70$                           \\ \hline
Pirouette                                                     & $\phi_{TTA} \ge 73$                                            & $73 \ge \phi_{TTA} \ge 68$     \\ \hline
\end{tabular}
\label{tab:perfLevels_nominal}
\end{table}

\begin{table}[]
\centering
\caption{Performance levels for aggressive maneuvering mission}
\begin{tabular}{|c|c|c|}
\hline
Maneuver                                                      & \begin{tabular}[c]{@{}c@{}}Desired TTA Score \\ (Level 1)\end{tabular} & \begin{tabular}[c]{@{}c@{}}Adequate TTA Score \\ (Level 2)\end{tabular} \\ \hline
\begin{tabular}[c]{@{}c@{}}Lateral Reposition\end{tabular} & $\phi_{TTA} \ge 82$                                             & $82 \ge \phi_{TTA} \ge 77$                           \\ \hline
Depart Abort                                                  & $\phi_{TTA} \ge 82$                                             & $82 \ge \phi_{TTA} \ge 77$                           \\ \hline
Pirouette                                                     & $\phi_{TTA} \ge 80$                                             & $80 \ge \phi_{TTA} \ge 75$      \\ \hline                     
\end{tabular}
\label{tab:perfLevels_aggressive}
\end{table}

\begin{figure}[h!]
  \centering
  \includegraphics[width=0.9\textwidth]{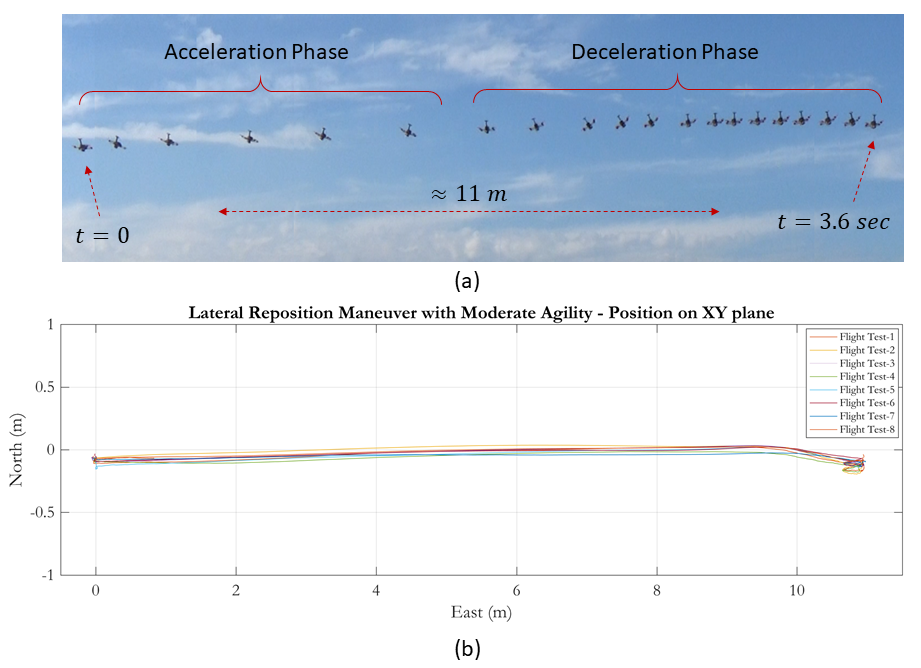}
  \caption{Lateral reposition flight test \cite{yuksek2020system}.}
  \label{fig:latRepo_test}
\end{figure}

\begin{figure}[h!]
  \centering
  \includegraphics[width=\textwidth]{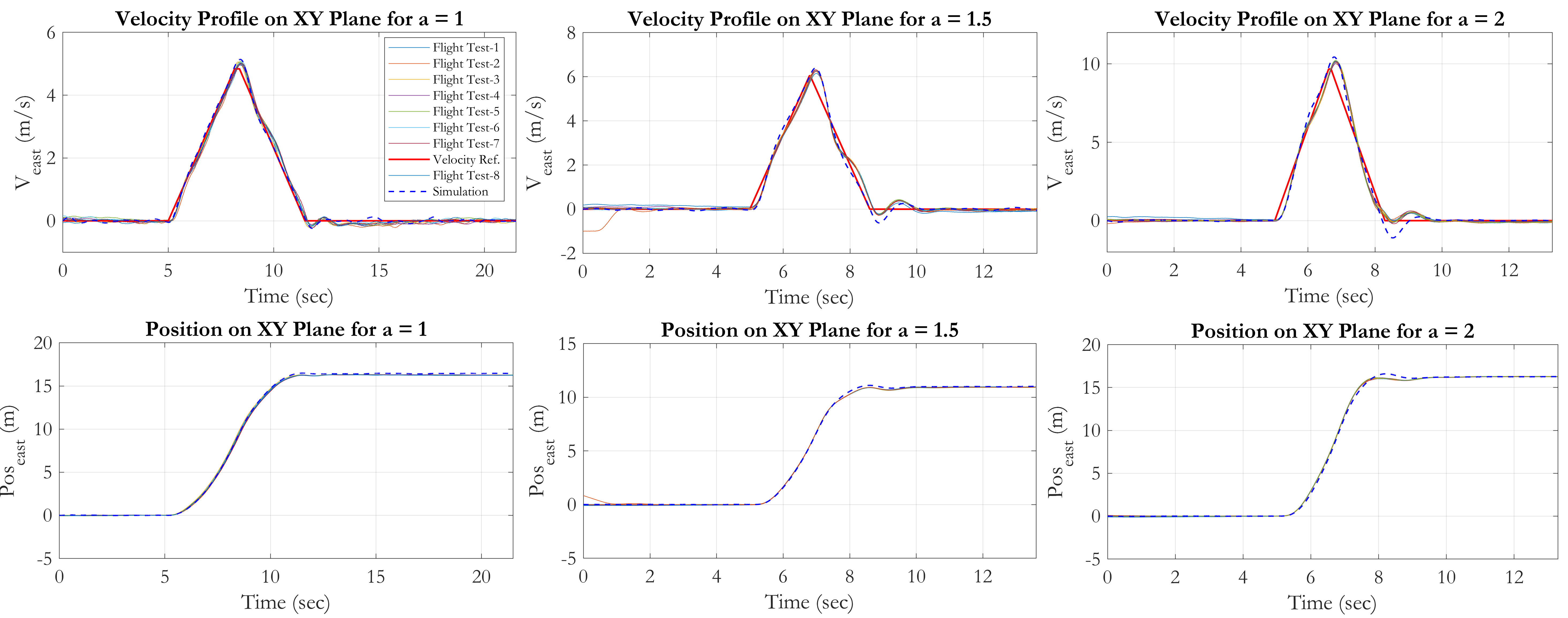}
  \caption{Velocity and position time histories for scaled-down lateral reposition MTE \cite{yuksek2020system}.}
  \label{fig:latRepo_timeHistory}
\end{figure}

A descriptive example for scaled-down lateral reposition MTE is given in Figure \ref{fig:latRepo_test} \cite{yuksek2020system}. In this test, a multirotor UAV is supposed to perform a lateral reposition manoeuvre while following a triangular-shape velocity profile. Velocity time histories for different agility levels are given in Figure \ref{fig:latRepo_timeHistory}.

\begin{figure}[h!]
  \centering
  \includegraphics[width=0.7\textwidth]{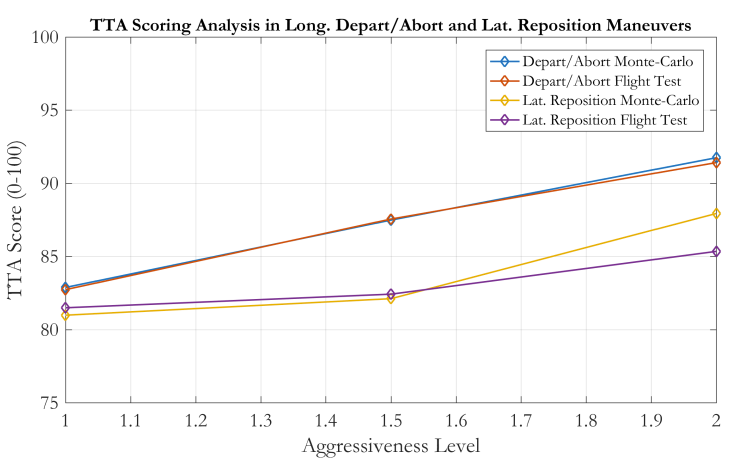}
  \caption{TTA analysis results for different aggressiveness levels \cite{yuksek2020system}.}
  \label{fig:tta_analysis}
\end{figure}

Upon completing simulations and flight tests, TTA score analysis is performed to evaluate the performance of the overall closed-loop system; the results are presented in Figure \ref{fig:tta_analysis}. Here, both Monte Carlo studies in the simulation environment and flight tests indicate that TTA scores for lateral reposition and depart/abort maneuvers meet the Level-1 TTA requirements which are given in Table \ref{tab:perfLevels_nominal} and \ref{tab:perfLevels_aggressive}.

In summary, control system specifications for autonomous systems are highly critical requirements which are directly connected with operation safety of the vehicles. In operation environments such as urban areas, farms or even in indoor environments, it is crucial to follow given reference signals (i.e. reference trajectories for outer-loop and reference attitude for inner-loop) to perform the given mission safely. If the vehicle could not track the given signals properly, this might result in catastrophic accidents and affects the system trust, directly. For these reasons, autonomous system specifications, especially for aerial vehicles, should be investigated for two levels, a) low-level attitude control loop, and b) trajectory tracking loop. Here, the former one defines attitude command tracking requirements in time-domain such as settling-time, rise-time and overshoot. An example is given which is applied on attitude control system design process of a satellite system. A similar approach could be applied for autonomous aerial and ground systems. The latter one defines specifications based on aggressiveness level and signal tracking error for trajectory tracking loop. A descriptive example is also given in which flight tests are performed to evaluate the trajectory tracking performance of a UAV. Here, it is shown that designed inner- and outer-loops meet the trajectory tracking and aggressiveness requirements which are defined by scaling-down the rotorcraft specifications. Results are validated in simulation environment and flight tests. For future works, definitions of autonomous system specifications could be defined for higher levels of the loop such as mission planning algorithms which contain trajectory generation and task assignment processes.

\section{Conclusion}
As autonomous systems become more widely adopted, the need for specifying their behaviour will become ever more pressing. A lack of specification makes it difficult or impossible to judge whether a system is functioning as intended or if there are deficiencies in its design. Likewise, imprecise specifications create difficulties from a regulatory standpoint; for instance, if the system conforms to an \emph{interpretation} of a certain imprecise requirement which is later found to be at variance with the interpretation by the regulatory body. Rigorous specifications, on the other hand, should leave no scope for ambiguity and hold the promise of enabling truly trustworthy designs; the difficulties arise in the effort that is currently required to describe the intended behaviour precisely and in the inherent complexity of reasoning about this behaviour. At present, formal specifications for autonomous systems cannot be said to be widely adopted by industry or government regulatory bodies, but this state of affairs may change in future if some of the difficulties inherent in formal specification and verification surveyed in this article can be even partially overcome.

\paragraph{Acknowledgements:} The authors would like to thank Dr Cora-Lisa Perner at Airbus, Dr Wilfried Steiner at TTTech Labs and Dr Oscar Gonzalez Villarreal at the School of Aerospace, Transport and Manufacturing, Cranfield University for their insights and interesting discussions. This work is supported by the UKRI Engineering and Physical Sciences Research Council [EPSRC Grant number: EP/V026763/1]

\bibliographystyle{alpha}
\bibliography{root.bib}

\end{document}